\documentclass[preprint,preprintnumbers,showpacs,aps,amssymb]{revtex4}

\usepackage{graphicx}
\usepackage{bm}
\usepackage{amsmath}

\def\calB{{\cal B}}

\def\bbar{{\bar b}}
\def\hbar{{\bar h}}

\def\qbar{{\bar q}}

\def\ubar{{\bar u}}
\def\vbar{{\bar v}}

\def\kslash{k\hspace{-1.8mm}/}

\def\gv{g_\perp^{(v)}}
\def\ga{g_\perp^{(a)}}

\def\nn{\nonumber}

\begin{document}
\title{Radiative $B\to K_1$ decays in the light-cone sum rules}
\author{Jong-Phil Lee}
\email{jplee@kias.re.kr}
\affiliation{Korea Institute for Advanced Study, Seoul 130-722, Korea}
\preprint{KIAS-P06035}

\begin{abstract}
The weak form factor for $B\to K_{1B}$ where $K_{1B}$ is the $^1P_1$ state
is calculated in the light-cone sum rules (LCSR).
Combining the quark model result for the form factor of $B\to K_{1A}$ with
$K_{1A}$ being the $^3P_1$ state, we have larger values for
$B\to K_1$ form factors than the previous LCSR results.
The increased form factors reduce the discrepancy between theory and the
experimental data for $B\to K_1 \gamma$.
Some phenomenological meanings are also discussed.
\end{abstract}

\maketitle
\section{Introduction}
Radiative $B$ decays to $K$ are a rich laboratory for the standard model and
new physics.
Especially, $B\to K^*\gamma$ is well understood theoretically via
$b\to s\gamma$ transition as well as experimentally.
Recently, higher resonant kaons are observed by CLEO and $B$ factories
\cite{CLEO}.
For example, BELLE collaboration has measured the radiative $B\to K_1$ decays
for the first time \cite{Abe:2004kr}:
\begin{eqnarray}
\calB(B^+\to K_1^+(1270)\gamma)&=&(4.28\pm0.94\pm0.43)\times 10^{-5}~,\\\nn
\calB(B^+\to K_1^+(1400)\gamma)&<&1.44\times 10^{-5}~,
\label{BELLE}
\end{eqnarray}
where $K_1$ is the orbitally excited axial vector meson.
In the theoretical side, recent developments of the QCD factorization (QCDF)
\cite{BBNS}
makes it possible to calculate the hard spectator contributions systematically
in a factorized form through the convolution at the heavy quark limit.
$B\to K^*\gamma$ is already studied in this line
\cite{Beneke:2000wa,Beneke:2001at,Bosch:2001gv,Ali}.
One good point about $K_1$ is that there are lots of things shared with
$B\to K^*\gamma$.
Basically both of them are governed by $b\to s\gamma$.
And the distribution amplitudes (DA) of $K^*$ and $K_1$ are same except the
overall factor of $\gamma_5$ which makes few differences in many calculations.
\par
A straightforward extension of the analysis for $B\to K^*\gamma$ to
$B\to K_1\gamma$ was given in \cite{jplee1}.
But the BELLE measurements of Eq.\ (\ref{BELLE}) reveal that theory predicts
much smaller branching ratio than data \cite{jplee2,Jamil}.
This is an opposite situation to that of $B\to K^*\gamma$ where theory predicts
larger branching ratio.
Considering the resemblance between $K^*$ and $K_1$, it is quite unlikely that
the same theoretical framework would produce discrepancies with experiment
in a reversed way.
\par
In the previous analysis the main uncertainty of theory lies in the
nonperturbative form factors.
Ref.\ \cite{jplee1} relies on the light-cone sum rule (LCSR) results for the
$B\to K_1$ form factors \cite{Safir}.
In \cite{Safir} only the leading twist DAs are considered without any
non-asymptotic contributions.
In this paper we revisit the $B\to K_1$ form factors in the LCSR.
There are three improvements compared to \cite{Safir}.
First, higher twist DAs are included; second, non-asymptotic contributions are
also considered; third, terms proportional to $m_A^2$, where $m_A$ is the mass
of axial meson, are not neglected.
\par
For $B\to K^*$ form factors, the LCSR results are updated \cite{Ball:2004rg}
up to the one-loop corrections to twist-2,3 contributions and leading order
twist-4.
It is thus legitimate to improve the theoretical accuracy for $B\to K_1$ form
factors.
\par
It is believed that the physical $K_1(1270)$ and $K_1(1400)$ states are the
mixtures of angular momentum eigenstates $^1P_1~(K_{1B})$ and $^3P_1~(K_{1A})$.
The mixing angle is not known precisely, but is close to the maximal.
This is a very natural and convenient way to explain the suppression of one
decay mode compared to the other.
For the suggestive angles $\theta=\pm37^\circ~,\pm58^\circ$ \cite{Cheng:2004yj},
negative ones are disfavored by (\ref{BELLE}).
\par
In \cite{Yang:2005gk}, some of the Gegenbauer moments of $K_{1B}$ DAs are
calculated by LCSR.
With this information, we explicitly calculate the $B\to K_{1B}$ form factor in
LCSR.
Since $K_{1B}$ and $K_{1A}$ have different G-parity, their Gegenbauer expansion
will not be the same.
Future study on $K_{1A}$ is necessary to reinforce the reliability of current
work.
We use the results from model calculations for $K_{1A}$ to give $B\to K_1$ form
factors.
This form factor will also be available for nonleptonic decay modes
\cite{Nardulli}
\par
The paper is organized as follows.
In the next section the weak form factors and axial vector meson DAs are
defined.
The LCSR evaluation is given in Sec. III.
Section IV deals with the LCSR results.
In Sec. V, some discussions about the results and their meanings appear.
Conclusions are also added at the end of this section.

\section{Form factors and distribution amplitudes}
For the axial vector $A(p_A,\epsilon)$, where $p_A$ ($\epsilon$) is the
momentum (polarization) of $A$, the relevant $B\to A$ transition matrix
elements are defined as \cite{Ebert:2001en,jplee1}
\begin{eqnarray}
\lefteqn{
\langle A(p_A,\epsilon)|\qbar i\sigma_{\mu\nu}q^\nu b|B(p_B)\rangle}\\\nn
&=&F_+^A(q^2)\left[(\epsilon^*\cdot q)(p_B +p_A)_\mu
 -\epsilon^*_\mu(m_B^2-m_A^2)\right]
 +F_-^A(q^2)\left[(\epsilon^*\cdot q) q_\mu-\epsilon^*_\mu q^2\right]\\\nn
&& +\frac{F_0^A(q^2)}{m_Bm_A}(\epsilon^*\cdot q)\left[
  (m_B^2-m_A^2)q_\mu-(p_B+p_A)q^2\right]~,\\
\lefteqn{
\langle A(p_A,\epsilon)|\qbar i\sigma_{\mu\nu}\gamma_5q^\nu b|B(p_B)\rangle
=-iF_+^A(q^2)\epsilon_{\mu\nu\alpha\beta}\epsilon^{*\nu}q^\alpha(p_A+p_B)^\beta
~,}
\end{eqnarray}
where $q=p_B-p_A$ and $m_B(m_A)$ is the $B$ (axial vector) meson mass.
We use $\epsilon_{0123}=+1$.
\par
The distribution amplitudes (DA) of the axial vector meson are given by
\cite{Ball:1998sk,Ball:1998ff,Yang:2005gk}
\begin{eqnarray}
\label{DAA}
\lefteqn{
\langle A(p_A,\epsilon)|\qbar_1(y)\gamma_\mu\gamma_5q_2(x)|0\rangle}\\\nn
&=&if_Am_A\int_0^1 du~e^{i(upy+\ubar px)}\left\{
 \frac{\epsilon^*\cdot z}{p\cdot z}p^A_\mu\left[\phi_\parallel(u)-\gv(u)\right]
 +\epsilon^*_\mu\gv(u)\right.\\\nn
&&\left.
 +\frac{\epsilon^*\cdot z}{2(p\cdot z)^2}m_A^2z_\mu\left[-\phi_\parallel(u)
 +2\gv(u)-g_3(u)\right]\right\}~,\\
\label{DAV}
\lefteqn{
\langle A(p_A,\epsilon)|\qbar_1(y)\gamma_\mu q_2(x)|0\rangle
=-if_Am_A\epsilon_{\mu\nu\alpha\beta}\epsilon^{*\nu}p^\alpha z^\beta
 \int_0^1 du~e^{i(upy+\ubar px)}~\frac{1}{4}\ga(u)~,}\\
\label{DAT}
\lefteqn{
\langle A(p_A,\epsilon)|\qbar_1(y)\sigma_{\mu\nu}\gamma_5q_2(x)|0\rangle}\\\nn
&=&f_A^\perp\int_0^1du~e^{i(upy+\ubar px)}\left\{
 (\epsilon^*_\mu p^A_\nu-\epsilon^*_\nu p^A_\mu)\phi_\perp(u)\right.\\\nn
&&\left.
 +\frac{\epsilon^*\cdot z}{(p\cdot z)^2}m_A^2(p^A_\mu z_\nu-p^A_\nu z_\mu)
 \left[-\frac{1}{2}\phi_\perp(u)+h_\parallel^{(t)}(u)
 -\frac{1}{2}h_3(u)\right]\right.\\\nn
&&\left.
 +\frac{m_A^2}{2p\cdot z}(\epsilon^*_\mu z_\nu-\epsilon^*_\nu z_\mu)
 \left[-\phi_\perp(u)+h_3(u)\right]\right\}~,\\
\label{DAP}
\lefteqn{
\langle A(p_A,\epsilon)|\qbar_1(y)\gamma_5q_2(x)|0\rangle
=f_A^\perp m_A^2(\epsilon^*\cdot z)\int_0^1 du~e^{i(upy+\ubar px)}~
\frac{1}{2}h_\parallel^{(s)}(u)~.}
\end{eqnarray}
Here $z=y-x$ and
\begin{equation}
p^\mu=p_A^\mu-\frac{m_A^2 z_\mu}{2p_A\cdot z}~,
\end{equation}
is the light-like vector, and $\ubar=1-u$.
The DAs $\phi_\parallel$ (twist-2), $\gv$, $\ga$ (twist-3), and
$g_3$ (twist-4) are anti-symmetric under the change $u\to\ubar$ while
$\phi_\perp$ (twist-2), $h_\parallel^{(t)}$, $h_\parallel^{(s)}$ (twist-3),
and $h_3$ (twist-4) are symmetric in $SU(3)$ limit,
because of the G-parity.
Thus
\begin{equation}
\int_0^1 du f(u)=0~,~~~{\rm for}~f=\phi_\parallel,~\ga,~\gv,~g_3.
\end{equation}
The leading twist DAs are expanded with the Gegenbauer polynomials.
In general, we can expand
\begin{eqnarray}
\phi_\parallel(u)&=&6u\ubar\sum_{l=0}^\infty a_l^\parallel C_l^{3/2}(u-\ubar)~,
\\\nn
\phi_\perp(u)&=&6u\ubar\Bigg[
1+\sum_{l=0}^\infty a_l^\perp C_l^{3/2}(u-\ubar)\Bigg]~.
\end{eqnarray}
For twist-3 DAs, the Wandzura-Wilczek type approximation will be used;
\begin{eqnarray}
\gv(u)&\simeq&\frac{1}{2}\Bigg[\int_0^udv~\frac{\phi_\parallel(v)}{\vbar}+
 \int_u^1dv~\frac{\phi_\parallel(v)}{v}\Bigg]~,\\\nn
\ga(u)&\simeq&2\Bigg[\ubar\int_0^udv~\frac{\phi_\parallel(v)}{\vbar}
 +u\int_u^1dv~\frac{\phi_\parallel(v)}{v}\Bigg]~,\\\nn
h_\parallel^{(t)}&\simeq&(u-\ubar)\Bigg[
 \int_0^udv~\frac{\phi_\perp(v)}{\vbar}-
 \int_u^1dv~\frac{\phi_\perp(v)}{v}\Bigg]~,\\\nn
h_\parallel^{(s)}&\simeq&2\Bigg[\ubar\int_0^udv~\frac{\phi_\perp(v)}{\vbar}
 +u\int_u^1dv~\frac{\phi_\perp(v)}{v}\Bigg]~.
\end{eqnarray}
The twist-4 DAs will not be considered afterwards.
The first few Gegenbauer coefficients are recently calculated by QCD sum rules
\cite{Yang:2005gk}.
\section{sum rule evaluation}

The main point of LCSR is to evaluate the two point correlation function:
\begin{equation}
\Pi_A=i\int d^4x~e^{-ip_B\cdot x}\langle A(p_A,\epsilon)|
 T\left[J(0)j_B^\dagger(x)\right]|0\rangle~.
\end{equation}
Here $j_B^\dagger(x)=\bbar(x)i\gamma_5 q(x)$ is the interpolating current
for $B$ meson, and $J(y)=\qbar(y)\Gamma b(y)$ is the heavy-to-light current
with $\Gamma$ being an appropriate gamma matrices.
To establish the sum rule, one calculates $\Pi_A$ in two ways.
On one hand, $\Pi_A$ is described in terms of hadronic observables.
We call this $\Pi_A^{\rm had}$.
Explicitly,
\begin{equation}
\Pi_A^{\rm had}=\frac{\langle A|J(0)|B\rangle
 \langle B|j_B^\dagger(0)|0\rangle}{m_B^2-p_B^2+i\epsilon}+({\rm res.})~,
\label{had}
\end{equation}
where the first term is the $B$ meson contribution and (res.) is the higher
resonance one.
The term $\langle A|J(0)|B\rangle$ defines the transition form factor while
\begin{equation}
\langle B|\bbar i\gamma_5 q|0\rangle
=\frac{m_B^2}{m_b}f_B~,
\end{equation}
is proportional to the $B$ meson decay constant $f_B$.
Here $\Pi_A^{\rm had}$ is considered as an analytic function of $p_B^2$.
Using the dispersion relation,
\begin{equation}
\Pi_A^{\rm had}=\int_{m_b^2}^\infty ds~\frac{\rho^{\rm had}(s)}{s-p_B^2}~,
\end{equation}
where $\rho^{\rm had}(s)$ is the spectral density function.
This is another expression of Eq.\ (\ref{had}), from which we can extract the
form of $\rho^{\rm had}$.
\par
On the other hand, $\Pi_A$ can be written by quarks and gluons, and hence by
light-cone distribution amplitudes (LCDAs).
We call this $\Pi^{\rm LC}$.
From the dispersion relation,
\begin{eqnarray}
\Pi_A^{\rm LC}&=&\int_{m_b^2}^\infty ds~\frac{\rho^{\rm LC}(s)}{s-p_B^2}\\\nn
&=&
\frac{1}{\pi}\int_{m_b^2}^\infty ds~\frac{{\rm Im}\Pi_A^{\rm LC}(s)}{s-p_B^2}~,
\end{eqnarray}
where the imaginary part of $\Pi_A^{\rm LC}$ will be expressed by the LCDAs.
At this stage, one assumes the quark-hadron duality for (res.) in
Eq.\ (\ref{had}) as
\begin{equation}
({\rm res.})=\frac{1}{\pi}ds~\int_{s_0}^\infty
 \frac{{\rm Im}\Pi_A^{\rm LC}(s)}{s-p_B^2}~,
\end{equation}
up to possible subtractions.
Here $s_0$ is the continuum threshold from which higher multi-particle states
begin.
In the numerical analysis, $s_0$ is considered as a hadronic parameter.
\par
Combining all this, one arrives at
\begin{equation}
\frac{\langle A|J(0)|B\rangle\langle B|j_B^\dagger(0)|0\rangle}{m_B^2-p_B^2}
=\frac{1}{\pi}\int_{m_b^2}^{s_0}ds~\frac{{\rm Im}\Pi_A^{\rm LC}(s)}{s-p_B^2}~.
\end{equation}
After the Borel transformation over $p_B^2$, we have the final expression
for the sum rule:
\begin{equation}
e^{-m_B^2/T}\langle A|J(0)|B\rangle\langle B|j_B^\dagger(0)|0\rangle
=\frac{1}{\pi}\int_{m_b^2}^{s_0}ds~e^{-s/T}~{\rm Im}\Pi_A^{\rm LC}(s)~,
\label{sumrule}
\end{equation}
where $T$ is the Borel parameter.
\par
Among the three form factors $F_{\pm,0}^A(q^2)$, the most important one is
$F_+^A(q^2=0)$ since only it is responsible for the radiative decay of
$B\to K_1$.
Also, it can be shown that $F_+^A(0)=F_-^A(0)$ \cite{Ball:2004rg}.
To extract $F_+^A$, we find it convenient to choose
$J(0)=\qbar i\sigma_{\mu\nu}\gamma_5q^\nu b$.
The left-hand-side (L.H.S.) of Eq.\ (\ref{sumrule}) is simply
\begin{equation}
({\rm L.H.S.})=-iF_+^A(q^2)\epsilon_{\mu\nu\alpha\beta}\epsilon^{*\nu}q^\alpha
(p_B+p_A)^\beta\left(\frac{m_B^2}{m_b}f_B\right)e^{-m_B^2/T}~.
\label{LHS}
\end{equation}
The right-hand-side (R.H.S.) of Eq.\ (\ref{sumrule}) is rather involved.
After contracting the $b\bbar$ quarks,
\begin{eqnarray}
({\rm R.H.S.})&=&
\frac{1}{\pi}\int_{m_b^2}^{s_0}ds~e^{-s/T}~{\rm Im}\int d^4x
\int\frac{d^4k}{(2\pi)^4}~\frac{e^{i(k-p_B)\cdot x}}{k^2-m_b^2+i\epsilon}\\\nn
&&\times
\Big[-\langle A|\qbar(0)\sigma_{\mu\nu}q^\nu\kslash q(x)|0\rangle
+m_b\langle A|\qbar(0)\sigma_{\mu\nu}q^\nu q(x)|0\rangle\Big]~.
\end{eqnarray}
The two matrix elements in the above equation can be written, after some gamma
matrix algebra, in terms of the LCDAs, Eqs.\ (\ref{DAA})-(\ref{DAP}).
In Eqs.\ (\ref{DAA})-(\ref{DAP}), the position coordinate $x$ can be replaced
effectively by
\begin{equation}
x^\mu\to\frac{\partial}{i\partial(\ubar p)_\mu}~,
\end{equation}
which is guaranteed by the presence of $e^{i\ubar px}$.
On the other hand, for the factor $1/p\cdot x$
\begin{eqnarray}
\frac{1}{p\cdot x}\phi(u)&\to& i\int_0^u dv~\phi(v)~,\\\nn
\frac{1}{(p\cdot x)^2} \phi(u)&\to& i^2\int_0^u dv\int_0^v dw~\phi(w)~,~~~
(\phi=\phi_\perp,~\ga~\gv,~g_3)
\end{eqnarray}
where the surface terms are vanishing.
In this way, one can remove $x$-dependence in (R.H.S.) except in the exponent.
Thus the integration over $x$ yields a delta function,
$\sim\delta^4(k-p_B+\ubar p)$.
Another delta function appears in the imaginary part of
$1/(k^2-m_B^2+i\epsilon)$.
Combining all together, one arrives at
\begin{eqnarray}
\lefteqn{
({\rm R.H.S.})}\\\nn
&=&
-i\epsilon_{\mu\nu\alpha\beta}\epsilon^{*\nu}q^\alpha p_A^\beta
\int_{m_b^2}^{s_0}ds~e^{-s/T}\int_0^1du\Bigg\{
f_Am_A\frac{1}{4}\ga(u)\Big[-2\delta_s+(-us+um_A^2-(1+\ubar)q^2)\delta_s'\Big]
\\\nn
&&
-f_Am_A\Big[\Phi_\parallel(u)\delta_s+u\gv(u)\delta_s-m_A^2G_3(u)\delta_s'\Big]
+f_A^\perp m_b\Big[\phi_\perp(u)\delta_s-m_A^2H_3(u)\delta_s'\Big]\Bigg\}~.
\label{RHS}
\end{eqnarray}
Here we use the short-hand notation,
$\delta_s\equiv\delta(s-m_b^2-2\ubar p\cdot p_B)$, and the differentiation
is with respect to $\ubar p$.
It is understood that at the final stage of calculation, $p\to p_A$.
And the newly defined functions are
\begin{eqnarray}
\Phi_\parallel(u)&\equiv&\int_0^u dv\Big[\phi_\parallel(v)-\gv(v)\Big]~,\\\nn
G_3(u)&\equiv&
\int_0^u dv\int_0^v dw \Big[-\phi_\parallel(w)+2\gv(w)-g_3(w)\Big]~,\\\nn
H_3(u)&\equiv&\int_0^u dv\Big[h_3(v)-\phi_\perp(v)\Big]~.
\end{eqnarray}
Equating Eqs.\ (\ref{LHS}) and (\ref{RHS}), after a little algebra, we have
the final expression for the form factor $F_+^A(q^2=0)$
\begin{eqnarray}
F_+^A(0)&=&
\frac{1}{2}e^{m_B^2/T}\left(\frac{m_b}{m_B^2f_B}\right)\Bigg\{
 f_Am_A\frac{e^{-s_0/T}}{s_0+m_A^2}\left[-\frac{s_0-m_A^2}{4}\ga(u_0)
 +m_A^2\frac{G_3(u_0)}{u_0}\right]\\\nn
&&
+f_Am_A\int_{u_0}^1\frac{du}{u}\exp\left[-\frac{m_b^2+\ubar m_A^2}{uT}\right]
\\\nn
&&
\times
 \left[-\frac{uT+m_b^2+(1-2u)m_A^2}{4uT}\ga(u)
-\Phi_\parallel(u)-u\gv(u)
 +\frac{m_A^2}{T}\frac{G_3(u)}{u}\right]\\\nn
&&
-f_A^\perp m_b e^{-s_0/T}\frac{m_A^2}{s_0+m_A^2}\frac{H_3(u_0)}{u_0}\\\nn
&&
 +f_A^\perp m_b\int_{u_0}^1\frac{du}{u}
 \exp\left[-\frac{m_b^2+\ubar m_A^2}{uT}\right]
 \left[\phi_\perp(u)-\frac{m_A^2}{T}\frac{H_3(u)}{u}\right]\Bigg\}~,
\label{finalSR}
\end{eqnarray}
where
\begin{equation}
u_0\equiv \frac{m_b^2+m_A^2}{s_0+m_A^2}~.
\end{equation}

\section{Results}

In what follows, only the case where $q^2=0$ is considered.
The basic input constants are summarized in Table \ref{input}.
\begin{table}
\begin{tabular}{cc||cc}\hline
\multicolumn{2}{c||}{hadronic information (in GeV)} &
\multicolumn{2}{c}{Gegenbauer moments (at 1 GeV)}\\ \cline{1-2}\cline{3-4}
$m_B$ & $5.27$  & $~~~~~~a^\parallel_0$ & $0.26$\\
$m_b$ & $4.8$   & $~~~~~~a^\parallel_1$ & $-1.75$\\
$f_B$ & $0.161$ & $~~~~~~a^\parallel_2$ & $0.13$\\
$m_A$ & $1.370$ & $~~~~~~a^\perp_1$ & $-0.13$\\
$f_A$ & $0.195$ & $~~~~~~a^\perp_2$ & $-0.02$\\
      &         & $~~~~~~a^\perp_3$ & $-0.02$
\label{input}
\end{tabular}
\caption{Input values. Gegenbauer moments are from \cite{Yang:2005gk}.}
\end{table}
The LCSR involves two important parameters, the continuum threshold $s_0$ and
the Borel parameter $T$.
Naively thinking, the continuum threshold is roughly
\begin{equation}
s_0\simeq(m_B+{\bar\Lambda})^2=(2m_B-m_b)^2~,
\end{equation}
where ${\bar\Lambda}\equiv m_B-m_b$.
Numerically,
\begin{equation}
s_0\simeq(2m_B-m_b)^2\equiv s_*\approx 33~{\rm GeV}^2~,
\end{equation}
for $m_B=5.27$ GeV and $m_b=4.8$ GeV is consistent with literatures
\cite{Ball:2004rg}.
We take this value as a starting point to fix $s_0$.
\par
In principle, $F_+^A$ is independent of the unphysical Borel parameter $T$.
But in reality there is a sum rule window of $T$ where a physical quantity is
stable.
If $T$ is too small, then the higher twist terms proportional to
$1/T^n~(n=1,2,\cdots)$ become too large.
One requires, for example,
\begin{equation}
\frac{\left(\frac{1}{T^n}~{\rm terms}\right)}
 {\left({\rm total}~F_+^A\right)}\lesssim 30\%~.
\label{lowerT}
\end{equation}
This condition imposes the lower bound of $T$.
The number $30\%$ might be changed, but we adopt this value here.
On the other hand, if $T$ is too large, then the contributions from the
continuum states become too large.
We require that
\begin{equation}
\frac{\frac{1}{\pi}\int_{s_0}^\infty ds~e^{-s/T}{\rm Im}\Pi(s)}
{\frac{1}{\pi}\int_{m_b^2}^\infty ds~e^{-s/T}{\rm Im}\Pi(s)}\lesssim 30\%~.
\label{upperT}
\end{equation}
This constraint imposes the upper bound of $T$.
Note that the condition of Eq.\ (\ref{upperT}) is used in \cite{Ball:2004rg}
to determine the lower bound of continuum threshold, $s_0$.
In this analysis, however, we start with $s_0=s_*$ to determine the sum rule
window, and then we fix $s_0$ from the best stability of $F_+^A$ within the
sum rule window.
\par
\begin{figure}
\includegraphics{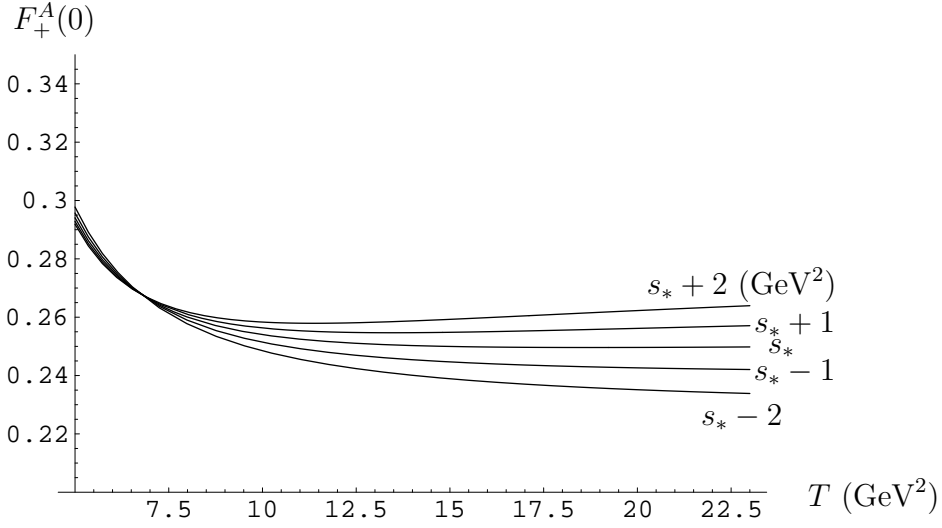}
\caption{$F_+^A(0)$ vs $T$ for various $s_0$ around $s_*$.}
\label{s*}
\end{figure}
\begin{figure}
\includegraphics{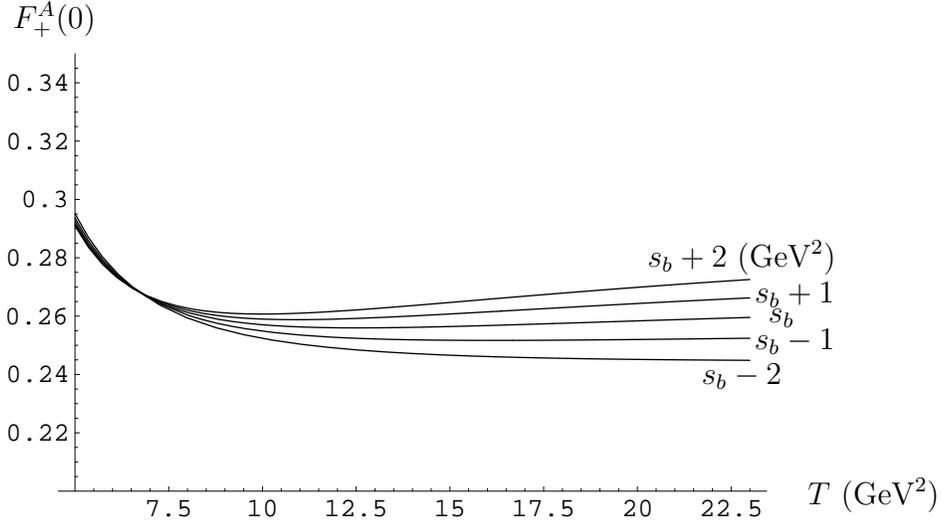}
\caption{$F_+^A(0)$ vs $T$ for various $s_0$ around $s_b$.}
\label{sb}
\end{figure}
\begin{figure}
\includegraphics{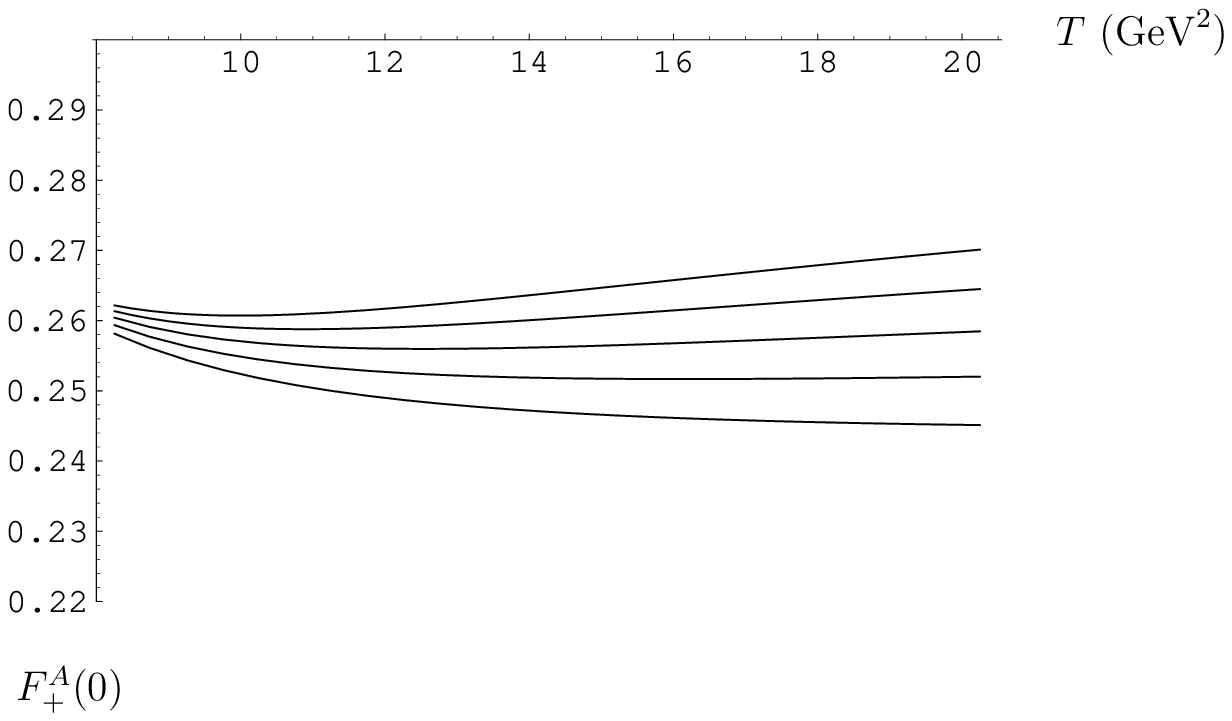}
\caption{A closer look of Fig.\ \ref{sb}.}
\label{sb2}
\end{figure}
\begin{figure}
\includegraphics{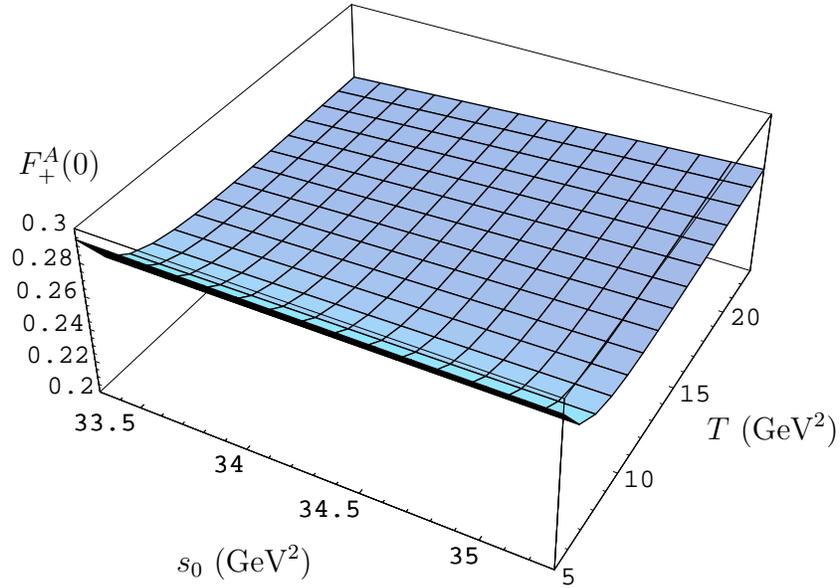}
\caption{3-dimensional plot of $F_+^A$.}
\label{3D}
\end{figure}

From Eqs.\ (\ref{lowerT}) and (\ref{upperT}) with $s_0=s_*$, we have
\begin{equation}
6.8\le T\le 21.7 ~({\rm GeV}^2)~.
\label{window}
\end{equation}
This window has overlaps with that of \cite{Safir}, but not with that
of \cite{Ball:2004rg} where only vector mesons are considered.
As an illustration, plots of $F_+^A$ over $T$ for various $s_0$ around $s_*$
are given in Fig.\ \ref{s*}.
To find the best value of $s_0$, we impose a simple condition.
We scan $s_0$ which minimize the value
$F_+^A(T_c+5~{\rm GeV}^2)-F_+^A(T_c-5~{\rm GeV}^2)$, where $T_c$ is the
central value of $T$ within the sum rule window.
We find that the best value of $s_0$ is
\begin{equation}
s_b\equiv 34.3~{\rm GeV}^2~.
\end{equation}
Plots of $F_+^A$ for various $s_0$ around $s_b$ are shown in Fig.\ \ref{sb}.
A closer look of Fig.\ \ref{sb} is given in Fig.\ \ref{sb2}, and 3-dimensional
plot of $F_+^A$ against $s_0$ and $T$ is given in Fig.\ \ref{3D}.
From the above analysis, we get
\begin{equation}
F_+^A(q^2=0;T_c)=0.256^{+0.0040}_{-0.0044}~,
\end{equation}
where the errors are from the variation of $s_0$ around $s_b$ by
$\pm1~{\rm GeV}^2$.
\par
The observed axial kaons $K_1(1270)$ and $K_1(1400)$ are mixtures of $^1P_1$
and $^3P_1$ states.
Their form factors are related via mixing angle $\theta$ as
\cite{Cheng:2004yj,PDG}
\begin{eqnarray}
F_i^{B\to K_1(1270)}&=&F_i^{A3}\sin\theta+F_i^A\cos\theta~,\\\nn
F_i^{B\to K_1(1400)}&=&F_i^{A3}\cos\theta-F_i^A\sin\theta~.
\end{eqnarray}
where $i=0,+,-$.
Here, $F_i^{A3}$ are the $^3P_1$ form factors.
We use the result of \cite{Cheng:2004yj}, $F_+^{A3}(q^2=0)=0.11$.
The mixing angle $\theta$ is not yet fixed precisely.
Ref.\ \cite{Cheng:2004yj} suggests $\theta=\pm 37^\circ,~\pm 58^\circ$.
Table \ref{mixing} shows the values of $F_+^{B\to K_1(1270)}$ and
$F_+^{B\to K_1(1400)}$ for these angles.
\begin{table}
\begin{tabular}{c||cccc}
$\theta$ & $37^\circ$ & $-37^\circ$ & $58^\circ$ & $-58^\circ$ \\\hline
$F_+^{B\to K_1(1270)}$ & $0.271$ & $0.138$ & $0.229$ & $0.042$\\
$F_+^{B\to K_1(1400)}$ & $-0.066$ & $0.242$ & $-0.159$ & $0.276$
\end{tabular}
\caption{$F_+^{B\to K_1(1270)}$ and $F_+^{B\to K_1(1400)}$ for various $\theta$.}
\label{mixing}
\end{table}
For negative angles, we find
\begin{equation}
F_+^{B\to K_1(1270)}<F_+^{B\to K_1(1400)}~.
\end{equation}
Since other parameters of the branching ratio are not so different in
$B\to K_1(1270)\gamma$ and $B\to K_1(1400)\gamma$, one expects
$\calB(B\to K_1(1270)\gamma)<\calB(B\to K_1(1400)\gamma)$ for the negative
mixing angles.
This is not consistent with the experimental data.

\section{Discussions and Conclusions}
As discussed in \cite{jplee2}, the discrepancy between theory and experiment for
$\calB(B\to K_1(1270,~1400)\gamma$ is mainly due to the smallness of the
relevant form factors.
If there is no mixing (i.e., $\theta=0$), then
$F_+^{B\to K_1(1270)}=F_+^A=0.256$.
This is considerably larger than the previous LCSR result of \cite{Safir},
$F_{+,{\rm Safir}}^{B\to K_1(1270)\gamma)}=0.14\pm0.03$.
The mixing effects are only $5.7\%$ and $-10.6\%$ for $\theta=37^\circ,~58^\circ$,
respectively.
In \cite{Safir}, only the asymptotic form of leading twist DA,
\begin{equation}
\phi^{\rm asy}_\perp(u)=6u(1-u)~,
\end{equation}
contributes to the sum rule.
According to Eq.\ (31) of \cite{Safir},
\begin{equation}
F_{+,{\rm Safir}}^A(0)
=\frac{1}{2}e^{m_B^2/T}\left(\frac{m_b}{m_B^2f_B}\right)f_A^\perp m_b
\int_{u_0}^1\frac{du}{u}\exp\left[-\frac{m_b^2+\ubar m_A^2}{uT}\right]
\phi_\perp^{\rm asy}(u)~.\
\label{Safir}
\end{equation}
It should be compared with Eq.\ (\ref{finalSR}).
Eq.\ (\ref{finalSR}) improves Eq.\ (\ref{Safir}) in three ways:
(1) higher twist DAs are included;
(2) non-asymptotic contributions are also included;
(3)
there is no term proportional to $m_A^2$ in $F_{+,{\rm Safir}}^A(0)$.
With the parameter set used in the previous section, we have
$F_{+,{\rm Safir}}^A(0;s=s_b;T=T_c)=0.187$.
This is lager than the value of
$F_{+,{\rm Safir}}^{B\to K_1(1270)\gamma)}=0.14\pm0.03$.
But if we take the sum rule window of Borel parameter adopted in \cite{Safir},
$F_{+,{\rm Safir}}^A(0;s=s_b;T=7.5~{\rm GeV}^2)=0.151$, which assures the
consistency of the present analysis.
We can check how much the new improvements contribute to the form factor.
The results are summarized in Table \ref{compare}.
\begin{table}
\begin{tabular}{c|ccc}
$F_{+,{\rm Safir}}^A(0)$ & $~~~~~~~+$(3) & $~~~~~~~+$(3)$+$(2)
& $~~~~~~~+$(3)$+$(2)$+$(1)\\\hline\hline
$0.187$ & $~~~~~~~0.237$ & $~~~~~~~0.155$ & $~~~~~~~0.256$
\end{tabular}
\caption{Contributions of new improvements. The parameter set used here is the
same as that in the previous section.
Conditions (1), (2), and (3) are explained in the text.}
\label{compare}
\end{table}
One finds that non-asymptotic and higher-twist DA contributions as well as
non-zero mass effects are considerable.
\par
The increase of the form factor will reduce the discrepancy between the
theoretical predictions and experimental data \cite{jplee2}.
At next-to-leading order of $\alpha_s$, the branching ratio of $B\to K_1\gamma$
is given by \cite{jplee1,jplee2}
\begin{equation}
\calB(B^0\to K_1^0\gamma)=0.003\left(1-\frac{m_{K_1}^2}{m_B^2}\right)^3\Big|
F_+^A(0)(-0.385-i0.014)+(f_A^\perp/{\rm GeV})(-0.024-i0.022)\Big|^2~.
\label{br}
\end{equation}
The resulting branching ratios are given in Table \ref{BR}.
\begin{table}
\begin{tabular}{c|cccc}
$\theta$ &
$~~~~~~~0^\circ$ & $~~~~~~~37^\circ$ & $~~~~~~~58^\circ$ &
$~~~~~~~$experiment \cite{Abe:2004kr}\\\hline\hline
$\calB(B^0\to K_1(1270)^0\gamma)$ &
$~~~~~~~2.81$ & $~~~~~~~3.14$ & $~~~~~~~2.27$ & $~~~~~~~4.28$\\
$\calB(B^0\to K_1(1400)^0\gamma)$ &
$~~~~~~~0.52$ & $~~~~~~~0.14$ & $~~~~~~~0.91$ & $~~~~~~~<1.44$
\end{tabular}
\caption{Branching ratios for various mixing angles in units of $10^{-5}$.}
\label{BR}
\end{table}
The enhancement is significant and the theoretical prediction becomes closer
to the experimental data compared to the previous analysis \cite{jplee1},
though there is still a gap.
\par
There are a few possibilities to improve further.
Firstly, the precisions are different between Eqs.\ (\ref{br}) and
(\ref{finalSR}).
Eq.\ (\ref{br}) contains the hard spectator interactions which appear as a
convolution between the jet function and the meson DAs.
The DAs contributing to Eq.\ (\ref{br}) are leading twist ones and of
asymptotic form.
It is thus necessary to include higher twist and non-asymptotic contributions
in Eq.\ (\ref{br}) at the same accuracy as was done in this work.
Also, Eq.\ (\ref{finalSR}) contains terms proportional to $m_A^2$, but
Eq.\ (\ref{br}) is the result of heavy quark limit.
One can easily expect that the next-to-leading order (NLO) of
$\Lambda_{\rm QCD}/m_b$ corrections to the QCDF framework might include the
terms of $m_A/m_b$, but there is no systematics so far.
The hard spectator interactions are given by the convolution of hard kernel
and meson DAs.
Similar non-zero $m_A^2$ terms will also appear in the axial vector DAs to
affect the hard spectator interactions.
But this effect is not expected to be large.
According to \cite{jplee1}, the hard spectator contributions amount to roughly
about $~5\%$ at the amplitude level.
\par
Secondly, $F_+^{A3}$ could be larger.
Actually there is no clue about the size of $F_+^{A3}$, but it might be that
$F_+^{A3}$ is comparable in size to $F_+^A$, just as in \cite{Cheng:2004yj}.
If this is the case, then the form factor can be enhanced via mixing
\begin{equation}
F_+^{B\to K_1(1270)}\approx 0.256\times(\sin\theta+\cos\theta)=0.35\sim0.36~,
\end{equation}
for $\theta=37^\circ,~58^\circ$, which results in a large branching ratio,
\begin{equation}
\calB(B^0\to K_1(1270)^0\gamma)\approx 5.3\times 10^{-5}~.
\end{equation}
\par
In conclusion, we have calculated $B\to K_{1B}$ form factor in LCSR.
This analysis improves the previous one in a few respects by including higher
twist DAs, non-asymptotic contributions, and non-zero $m_A^2$ terms.
The value is rather larger than the previous calculation and that from the
quark model result.
Larger value is well accommodated to the experimental data.
One needs more information about $K_{1A}$ and the mixing angle to reduce
theoretical uncertainties.
To go beyond the current work, one can include the NLO of $\alpha_s$
which might not be so different from that for $B\to K^*$ \cite{Ball:2004rg}.
And the study of $B\to K_1\gamma$ at higher accuracy comparable to this work
will be necessary.


\begin{thebibliography}{99}
\bibitem{CLEO}
  T.~E.~Coan {\it et al.}  [CLEO Collaboration],
  Phys.\ Rev.\ Lett.\  {\bf 84}, 5283 (2000)
  [arXiv:hep-ex/9912057];
S.~Nishida {\it et al.}  [Belle Collaboration],
  Phys.\ Rev.\ Lett.\  {\bf 89}, 231801 (2002)
  [arXiv:hep-ex/0205025];
B.~Aubert {\it et al.}  [BABAR Collaboration],
  Phys.\ Rev.\ D {\bf 70}, 091105 (2004)
  [arXiv:hep-ex/0409035].
\bibitem{Abe:2004kr}
  K.~Abe {\it et al.}  [BELLE Collaboration],
  arXiv:hep-ex/0408138;
  H.~Yang {\it et al.},
  Phys.\ Rev.\ Lett.\  {\bf 94}, 111802 (2005)
  [arXiv:hep-ex/0412039].
\bibitem{BBNS}
  M.~Beneke, G.~Buchalla, M.~Neubert and C.~T.~Sachrajda,
  Nucl.\ Phys.\ B {\bf 591}, 313 (2000)
  [arXiv:hep-ph/0006124].
\bibitem{Beneke:2000wa}
  M.~Beneke and T.~Feldmann,
  Nucl.\ Phys.\ B {\bf 592}, 3 (2001)
  [arXiv:hep-ph/0008255].
\bibitem{Beneke:2001at}
  M.~Beneke, T.~Feldmann and D.~Seidel,
  Nucl.\ Phys.\ B {\bf 612}, 25 (2001)
  [arXiv:hep-ph/0106067].
\bibitem{Bosch:2001gv}
  S.~W.~Bosch and G.~Buchalla,
  Nucl.\ Phys.\ B {\bf 621}, 459 (2002)
  [arXiv:hep-ph/0106081].
\bibitem{Ali}
  A.~Ali and A.~Y.~Parkhomenko,
  Eur.\ Phys.\ J.\ C {\bf 23}, 89 (2002)
  [arXiv:hep-ph/0105302].
\bibitem{jplee1}
J.~P.~Lee,
  Phys.\ Rev.\ D {\bf 69}, 114007 (2004)
  [arXiv:hep-ph/0403034].
\bibitem{jplee2}
Y.~J.~Kwon and J.~P.~Lee,
  Phys.\ Rev.\ D {\bf 71}, 014009 (2005)
  [arXiv:hep-ph/0409133].
\bibitem{Jamil}
  M.~Jamil Aslam and Riazuddin,
  Phys.\ Rev.\ D {\bf 72}, 094019 (2005)
  [arXiv:hep-ph/0509082];
M.~J.~Aslam,
  [arXiv:hep-ph/0604025].
\bibitem{Safir}
  A.~S.~Safir,
  Eur.\ Phys.\ J.\ directC {\bf 3}, 15 (2001)
  [arXiv:hep-ph/0109232].
\bibitem{Ball:2004rg}
  P.~Ball and R.~Zwicky,
  Phys.\ Rev.\ D {\bf 71}, 014029 (2005)
  [arXiv:hep-ph/0412079].
\bibitem{Cheng:2004yj}
  H.~Y.~Cheng and C.~K.~Chua,
  Phys.\ Rev.\ D {\bf 69}, 094007 (2004)
  [arXiv:hep-ph/0401141].
\bibitem{Yang:2005gk}
  K.~C.~Yang,
  JHEP {\bf 0510}, 108 (2005)
  [arXiv:hep-ph/0509337].
\bibitem{Nardulli}
  G.~Nardulli and T.~N.~Pham,
  Phys.\ Lett.\ B {\bf 623}, 65 (2005)
  [arXiv:hep-ph/0505048].
\bibitem{Ebert:2001en}
  D.~Ebert, R.~N.~Faustov, V.~O.~Galkin and H.~Toki,
  Phys.\ Rev.\ D {\bf 64}, 054001 (2001)
  [arXiv:hep-ph/0104264].
\bibitem{Ball:1998sk}
  P.~Ball, V.~M.~Braun, Y.~Koike and K.~Tanaka,
  Nucl.\ Phys.\ B {\bf 529}, 323 (1998)
  [arXiv:hep-ph/9802299].
\bibitem{Ball:1998ff}
  P.~Ball and V.~M.~Braun,
  Nucl.\ Phys.\ B {\bf 543}, 201 (1999)
  [arXiv:hep-ph/9810475].
\bibitem{PDG}
  S.~Eidelman {\it et al.}  [Particle Data Group],
  Phys.\ Lett.\ B {\bf 592} (2004) 1.
\end{thebibliography}
\end{document}